\documentclass[3p,times,procedia]{elsarticle}
\usepackage{nupha_ecrc}

\volume{00}

\firstpage{1}

\journalname{Nuclear Physics A}

\runauth{Jiaxing Zhao}

\jid{nupha}

\jnltitlelogo{Nuclear Physics A}

\usepackage{amssymb}

\usepackage[figuresright]{rotating}

\begin{document}

\begin{frontmatter}



\dochead{XXVIIIth International Conference on Ultrarelativistic Nucleus-Nucleus Collisions\\ (Quark Matter 2019)}

\title{Sequential hadronization in heavy ion collisions}


 \author[label1]{Jiaxing Zhao} \author[label2]{Shuzhe Shi}\author[label3]{Nu Xu}\author[label1]{Pengfei Zhuang}
  \address[label1]{Physics Department, Tsinghua University, Beijing 100084, China}
  \address[label2]{Department of Physics, McGill University, Montr\'eal, QC H3A 2T8, Canada}
  \address[label3]{ Institute of Modern Physics, Chinese Academy of Sciences, Lanzhou, Gansu 730000, China}
  

\address{}

\begin{abstract}
Heavy flavor supplies a chance to constrain and improve the hadronization mechanism. We have established a sequential coalescence model with charm conservation and applied it to the charmed hadron production in heavy ion collisions. The charm conservation enhances the earlier hadron production and suppresses the later production. This relative enhancement (suppression) changes significantly the ratios between charmed hadrons in heavy ion collisions. 
\end{abstract}


\end{frontmatter}


\   \\

The hadronization mechanism is a fundamental problem in QCD, and due to the running coupling constant the hadronization is a non-perturbative process, it's hard to deal with. Different from the hadronization process in the vacuum, such as fragmentation, the statistic hadronization plays an important role in quark hadronization from quark-gluon plasma. The yield of various hadrons can be described well by the thermal statistical model~\cite{BraunMunzinger:2003zd}. More dynamical approaches are the coalescence models that show more power to explain the light hadron properties in heavy ion collisions, especially the quark number scaling of the elliptic flow and the enhancement of the baryon to meson ratio~\cite{Molnar:2003ff,Greco:2003mm,Fries:2003vb}. 

However, there are still some questions need to be remarked:
1). Energy conservation and entropy. The kinematics of the coalescence process is $2\to1$ or $3\to1$, which makes it impossible to conserve 4-momentum~\cite{Fries:2008hs}. The quark coalescence model has been extended to include finite width that takes into account off-shell effects which allows to include the constraint of energy conservation~\cite{Ravagli:2007xx}. Besides, coalescence through instantaneous projection seems to reduce the number of particles. That raises the question of whether the entropy is conserved(increased) or not. Although entropy depends not only on the number of particles but also on the degeneracies in both phases as well as on the masses, it remains a challenge to find a consistent approach to conserving energy and conserving or increasing entropy, together with a good description of single-particle spectra and elliptic flow for both low and intermediate $p_T$. 2). Coalescence probability. The Wigner function in coalescence formula is treated as coalescence probability for $n$ quarks to combine into a hadron. In the previous studies, people use a Gaussian shape in space and momentum as coalescence probability with the width as a free parameter. The Wigner function can be constructed by the wavefunction of the hadrons self-consistently by:
\begin{eqnarray}
\label{td}
W(r,q)=\int d^4y e^{-ipy}\Psi(r+y/2)\Psi^*(r-y/2).
\end{eqnarray}
Unlike the light hadrons, the potential model can be used to describe the properties of heavy flavor hadrons in both vacuum and finite temperature~\cite{Satz:2005hx,Crater:2008rt,Zhao:2017gpq,Shi:2019tji}. And the wavefunction $\Psi(r)$ of heavy flavor hadrons can be obtained by solving two- and three-body Schr\"odinger equation or Dirac equation.
3). Heavy quark number conservation. Because the charm quark mass $m_Q$ is much larger than the typical temperature of the hot medium which formed in the heavy ion collisions at RHIC and LHC energies, $m_Q\gg T\sim 500$MeV, the charm quark number is almost contributed by the initial production and conserved during the whole evolution of the colliding system~\cite{Zhou:2016wbo}. The question is how to ensure the charm quark number conservation self-consistently in the coalescence mechanism? 
4). Hadronization sequence. Different from light hadrons which formed at confinement and de-confinement phase transition temperature $T_c$, heavy flavor hadrons can survive at a higher temperature which would be produced earlier. This hadronization sequence has been observed in the heavy quarkonium system~\cite{Chen:2013wmr, Du:2015wha}. And the recent experimental data  shows the $D^0$ and $\phi$ seem to decouple from the system earlier and gain less radial collectivity compared with light hadrons~\cite{Adam:2018inb}.

The hadronization of heavy flavor supply a chance to constrain and improve the coalescence mechanism. 
In our previous study~\cite{Zhao:2018jlw}, we have established a sequential coalescence model with charm conservation and applied it to the charmed hadron production in heavy ion collisions. The charm conservation effect can be realized self-consistently in the sequential coalescence model. Hadronization sequence and coalescence probability of charmed hadrons are determined by two- and three-body Dirac equation. 

\begin{figure}[!htb]
{$$\includegraphics[width=0.35\textwidth]{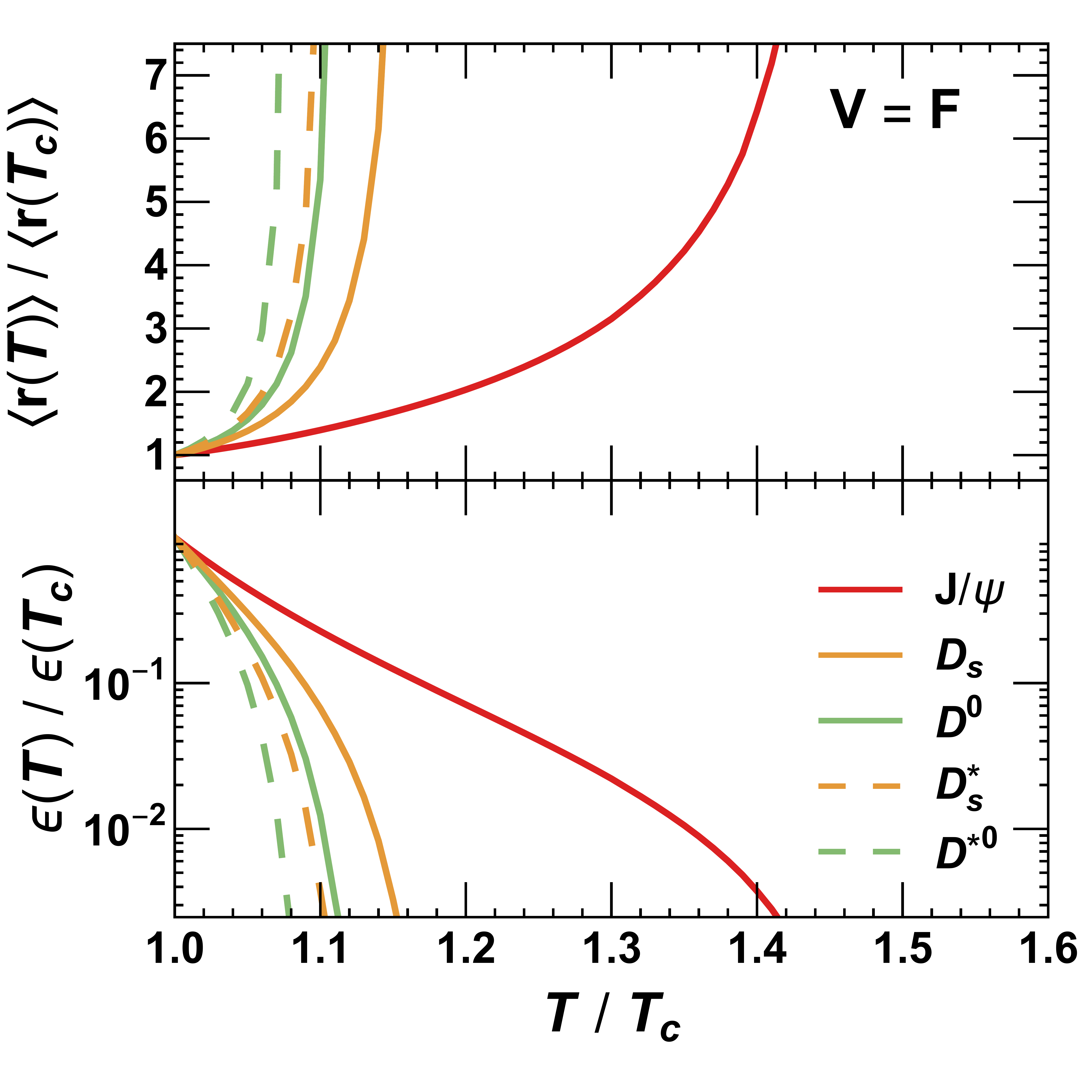} \   \    \    \includegraphics[width=0.35\textwidth]{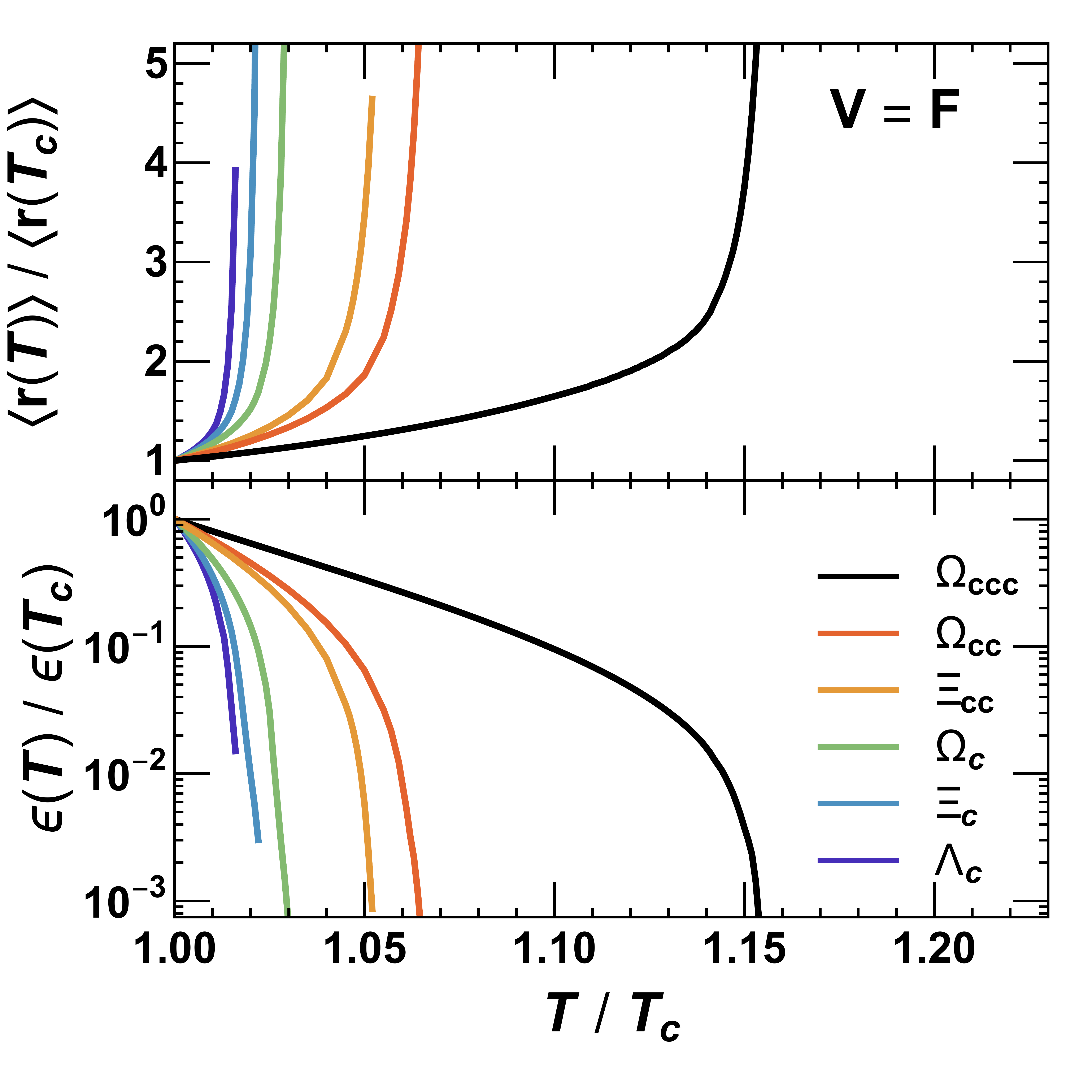}$$
\caption{The scaled charmed mesons(left panel), charmed baryons(right panel) binding energy and root-mean-squared radius as functions of temperature.}
\label{fig1}}
\end{figure}

In the framework of two- and three-body Dirac equations~\cite{Shi:2019tji}, we systematically study the static properties of heavy flavor hadrons in finite temperature. For a rapid dissociation where there is no heat exchange between the constituent quarks and the medium, the potential between quarks is the internal energy $U$, while for a slow dissociation, there is enough time for the constituent quarks to exchange heat with the medium, the free energy $F$ which is simulated by lattice QCD is treated as potential~\cite{Petreczky:2010yn}. In a general case in high energy nuclear collisions, the potential $V$ is in between the two limits, $|F|<|V|<|U|$. The averaged radius and binding energy of charmed hadrons with free energy $F$ are showed in Fig.\ref{fig1}. As the temperature increases, the color screening will enhance, so the averaged radius of charmed hadron increases and the binding energy drops down. The dissociation temperature $T_d$ is determined by vanishing the binding energy $\epsilon(T)\to 0$ or infinite the averaged radius $\langle r(T) \rangle \to \infty$. We find there exist an obvious dissociation sequence:  
\begin{eqnarray}
\label{td}
T_d^{J/\psi}>T_d^{D_s}>T_d^{\Omega_{ccc}}>T_d^{D^0}>T_d^{\Omega_{cc},\Xi_{cc}}>T_d^{\Omega_c, \Xi_c, \Lambda_{c}}>T_d^{\pi, K, N}\approx T_c
\end{eqnarray}

The evolution of the hot medium created in the heavy ion collisions can be described by the hydrodynamic equations $\partial_\mu T^{\mu\nu}=0$. During the evolution of the hot medium, the temperature continuously drops down due to the expansion of the system. When the local temperature $T({\bf x},\tau_h)$ smaller than the dissociation temperature $T_d^h$ of charmed hadron $h$, the corresponding hadrons $h$ will start to be formed. The dissociation sequence gives the production sequence.
The spectrum of charmed hadrons can be calculated with the coalescence formula,
\begin{eqnarray}
\label{coalescence}
\frac{dN_h}{d^2 P_Td\eta} &=& C \int P^\mu d\sigma_\mu \prod_{i=1}^n{\frac{d^4 x_i d^4 p_i}{(2\pi)^3}}f_i(x_i,p_i)\times W_h( x_1,...,x_i, p_1,...,p_i).
\end{eqnarray}
 Where the constant $C=(2J+1)/(3^n 2^n)$ is the statistical factor to take into account the internal quantum numbers. The integration is on the coalescence hypersurface $\sigma_\mu(\tau_h,{\bf X})$. The summation is over the constituent quarks with $n=2$ for mesons and $n=3$ for baryons. Where $W_h$ is the Wigner function which is constructed directly from the wavefunction. $f_i$ in the spectra (\ref{coalescence}) is the distribution function of the constituent quarks in phase space. The light quarks $u$ and $d$ are thermalized in the medium so their distribution is followed by the Fermi-Dirac distribution. Considering that strange quarks may not reach fully chemical equilibrium at RHIC energy, the fugacity factor $\gamma_s$ is included~\cite{Castorina:2017dvt}. The charm quarks are produced through initial hard processes and then interact with the hot medium. Considering the energy loss during the motion, the charm quark distribution $f_c(x,p)$ is controlled by a transport approach. In our previous work, we take, as a first approximation, a linear combination of $f_{pp}$ and thermal distribution $f_{th}$ as the charm quark distribution, $f_c(x,p) = r_h\rho_c(x|{\bf b})[\alpha f_{th}(p)+\beta f_{pp}(p)]$. The coefficients $\alpha$ and $\beta$ control the degree of thermalization of charm quarks, reflect the continuous thermalization charm quarks in hot medium. The space density $\rho_c(x|{\bf b})$ is the superposition of $pp$ collisions. 
Aiming to study charm conservation effect, one needs to include all charmed hadrons in the calculation.
The time-dependent charm quark number fraction $r_h(\tau)$ describes the charm conservation during the hadronization. Different from simultaneous production, more charm quarks are involved in the earlier hadronization, and fewer charm quarks join the later hadronization. The charm quark number fraction $r_h=1$ for $D_s$, $1-N_{D_s}/N_c\sim 0.9$ for the other charmed mesons, and $1-N_m/N_c\sim0.6$ for charmed baryons. 

\begin{figure}[!htb]
{$$\includegraphics[width=0.6\textwidth]{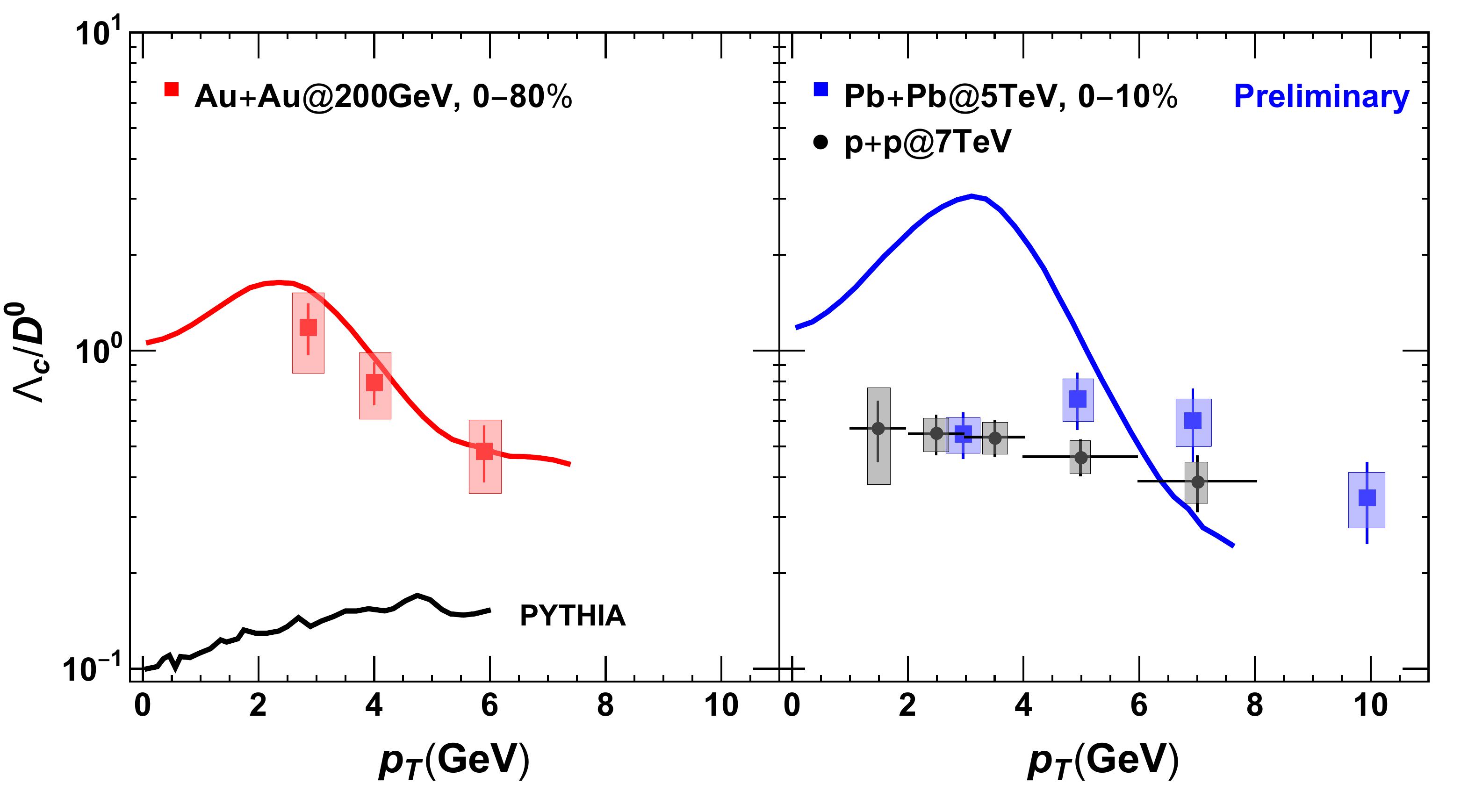}$$
\caption{The yield ratio of $\Lambda_c^+/D^0$ at RHIC and LHC energies. The experimental data in A+A and p+p collisions are from STAR~\cite{Adam:2019hpq} and ALICE~\cite{Acharya:2017kfy,Sirunyan:2019fnc} collaborations, the solid and dashed lines are respectively our sequential and simultaneous coalescence calculations.}
\label{fig2}}
\end{figure}

The precise measurement of $D_s$, $D^0$ and charm baryons in the experiment give us a chance to study the charm hadronization mechanism in heavy ion collisions. In our model, both the strangeness enhancement and the charm conservation are responsible for the $D_s/D^0$ ratio enhancement. The $D_s^+$ enhancement leads to a $D^0$ suppression and in turn to a further $D_s^+/D^0$ enhancement~\cite{Zhao:2018jlw}. The prediction explains well the $D_s/D^0$ ratio which observed in heavy ion collisions at RHIC. 
For the charmed baryons, considering the difference in statistics for two- and three-body states shown in the hadron spectra (\ref{coalescence}), the baryon to meson ratio in A+A collisions will be dramatically enhanced comparing with p+p collisions in coalescence models. Our prediction about the yield ratio of $\Lambda_c^+/D^0$ is agreed well with experimental data from STAR, as shown in Fig.~\ref{fig2}. However, the preliminary experimental data from ALICE shows a big difference from that in STAR and also the theoretical prediction. 

In summary, we have built a sequential coalescence mechanism which can consider the charm conservation effect self-consistently. We applied it to the charmed hadrons production in heavy ion collisions at RHIC and LHC. We found charm conservation leads to an enhancement for earlier produced hadrons and a suppression for later produced hadrons. In the future, we will consider the contribution from fragmentation and the space-momentum correlation in our sequential coalescence model.




\bibliographystyle{elsarticle-num}
\bibliography{refs}

\begin{thebibliography}{10}
\expandafter\ifx\csname url\endcsname\relax
  \def\url#1{\texttt{#1}}\fi
\expandafter\ifx\csname urlprefix\endcsname\relax\def\urlprefix{URL }\fi
\expandafter\ifx\csname href\endcsname\relax
  \def\href#1#2{#2} \def\path#1{#1}\fi

\bibitem{BraunMunzinger:2003zd}
P.~Braun-Munzinger, K.~Redlich, J.~Stachel, {Particle production in heavy ion
  collisions} (2003) 491--599\href {http://arxiv.org/abs/nucl-th/0304013}
  {\path{arXiv:nucl-th/0304013}}.

\bibitem{Molnar:2003ff}
D.~Molnar, S.~A. Voloshin, {Elliptic flow at large transverse momenta from
  quark coalescence}, Phys. Rev. Lett. 91 (2003) 092301.
\newblock \href {http://arxiv.org/abs/nucl-th/0302014}
  {\path{arXiv:nucl-th/0302014}}, \href
  {http://dx.doi.org/10.1103/PhysRevLett.91.092301}
  {\path{doi:10.1103/PhysRevLett.91.092301}}.

\bibitem{Greco:2003mm}
V.~Greco, C.~M. Ko, P.~Levai, {Parton coalescence at RHIC}, Phys. Rev. C68
  (2003) 034904.
\newblock \href {http://arxiv.org/abs/nucl-th/0305024}
  {\path{arXiv:nucl-th/0305024}}, \href
  {http://dx.doi.org/10.1103/PhysRevC.68.034904}
  {\path{doi:10.1103/PhysRevC.68.034904}}.

\bibitem{Fries:2003vb}
R.~J. Fries, B.~Muller, C.~Nonaka, S.~A. Bass, {Hadronization in heavy ion
  collisions: Recombination and fragmentation of partons}, Phys. Rev. Lett. 90
  (2003) 202303.
\newblock \href {http://arxiv.org/abs/nucl-th/0301087}
  {\path{arXiv:nucl-th/0301087}}, \href
  {http://dx.doi.org/10.1103/PhysRevLett.90.202303}
  {\path{doi:10.1103/PhysRevLett.90.202303}}.

\bibitem{Fries:2008hs}
R.~J. Fries, V.~Greco, P.~Sorensen, {Coalescence Models For Hadron Formation
  From Quark Gluon Plasma}, Ann. Rev. Nucl. Part. Sci. 58 (2008) 177--205.
\newblock \href {http://arxiv.org/abs/0807.4939} {\path{arXiv:0807.4939}},
  \href {http://dx.doi.org/10.1146/annurev.nucl.58.110707.171134}
  {\path{doi:10.1146/annurev.nucl.58.110707.171134}}.

\bibitem{Ravagli:2007xx}
L.~Ravagli, R.~Rapp, {Quark Coalescence based on a Transport Equation}, Phys.
  Lett. B655 (2007) 126--131.
\newblock \href {http://arxiv.org/abs/0705.0021} {\path{arXiv:0705.0021}},
  \href {http://dx.doi.org/10.1016/j.physletb.2007.07.043}
  {\path{doi:10.1016/j.physletb.2007.07.043}}.

\bibitem{Satz:2005hx}
H.~Satz, {Colour deconfinement and quarkonium binding}, J. Phys. G32 (2006)
  R25.
\newblock \href {http://arxiv.org/abs/hep-ph/0512217}
  {\path{arXiv:hep-ph/0512217}}, \href
  {http://dx.doi.org/10.1088/0954-3899/32/3/R01}
  {\path{doi:10.1088/0954-3899/32/3/R01}}.

\bibitem{Crater:2008rt}
H.~W. Crater, J.-H. Yoon, C.-Y. Wong, {Singularity Structures in Coulomb-Type
  Potentials in Two Body Dirac Equations of Constraint Dynamics}, Phys. Rev.
  D79 (2009) 034011.
\newblock \href {http://arxiv.org/abs/0811.0732} {\path{arXiv:0811.0732}},
  \href {http://dx.doi.org/10.1103/PhysRevD.79.034011}
  {\path{doi:10.1103/PhysRevD.79.034011}}.

\bibitem{Zhao:2017gpq}
J.~Zhao, P.~Zhuang, {Multicharmed Baryon Production in High Energy Nuclear
  Collisions}, Few Body Syst. 58~(2) (2017) 100.
\newblock \href {http://dx.doi.org/10.1007/s00601-017-1255-9}
  {\path{doi:10.1007/s00601-017-1255-9}}.

\bibitem{Shi:2019tji}
S.~Shi, J.~Zhao, P.~Zhuang, {Heavy Flavor Dissociation in the Frame of
  Multi-Body Dirac Equations}\href {http://arxiv.org/abs/1905.10627}
  {\path{arXiv:1905.10627}}.

\bibitem{Zhou:2016wbo}
K.~Zhou, Z.~Chen, C.~Greiner, P.~Zhuang, {Thermal Charm and Charmonium
  Production in Quark Gluon Plasma}, Phys. Lett. B758 (2016) 434--439.
\newblock \href {http://arxiv.org/abs/1602.01667} {\path{arXiv:1602.01667}},
  \href {http://dx.doi.org/10.1016/j.physletb.2016.05.051}
  {\path{doi:10.1016/j.physletb.2016.05.051}}.

\bibitem{Chen:2013wmr}
B.~Chen, Y.~Liu, K.~Zhou, P.~Zhuang, {$\psi^\prime$ Production and $B$ Decay in
  Heavy Ion Collisions at {LHC}}, Phys. Lett. B726 (2013) 725--728.
\newblock \href {http://arxiv.org/abs/1306.5032} {\path{arXiv:1306.5032}},
  \href {http://dx.doi.org/10.1016/j.physletb.2013.09.036}
  {\path{doi:10.1016/j.physletb.2013.09.036}}.

\bibitem{Du:2015wha}
X.~Du, R.~Rapp, {Sequential Regeneration of Charmonia in Heavy-Ion Collisions},
  Nucl. Phys. A943 (2015) 147--158.
\newblock \href {http://arxiv.org/abs/1504.00670} {\path{arXiv:1504.00670}},
  \href {http://dx.doi.org/10.1016/j.nuclphysa.2015.09.006}
  {\path{doi:10.1016/j.nuclphysa.2015.09.006}}.

\bibitem{Adam:2018inb}
J.~Adam, et~al., {Centrality and transverse momentum dependence of $D^0$-meson
  production at mid-rapidity in Au+Au collisions at ${\sqrt{s_{\rm NN}} =
  \rm{200\,GeV}}$}, Phys. Rev. C99~(3) (2019) 034908.
\newblock \href {http://arxiv.org/abs/1812.10224} {\path{arXiv:1812.10224}},
  \href {http://dx.doi.org/10.1103/PhysRevC.99.034908}
  {\path{doi:10.1103/PhysRevC.99.034908}}.

\bibitem{Zhao:2018jlw}
J.~Zhao, S.~Shi, N.~Xu, P.~Zhuang, {Sequential Coalescence with Charm
  Conservation in High Energy Nuclear Collisions}\href
  {http://arxiv.org/abs/1805.10858} {\path{arXiv:1805.10858}}.

\bibitem{Petreczky:2010yn}
P.~Petreczky, {Quarkonium in Hot Medium}, J. Phys. G37 (2010) 094009.
\newblock \href {http://arxiv.org/abs/1001.5284} {\path{arXiv:1001.5284}},
  \href {http://dx.doi.org/10.1088/0954-3899/37/9/094009}
  {\path{doi:10.1088/0954-3899/37/9/094009}}.

\bibitem{Castorina:2017dvt}
P.~Castorina, S.~Plumari, H.~Satz, {Strangeness Production and Color
  Deconfinement}, Int. J. Mod. Phys. E26~(12) (2017) 1750081.
\newblock \href {http://arxiv.org/abs/1709.02706} {\path{arXiv:1709.02706}},
  \href {http://dx.doi.org/10.1142/S0218301317500811}
  {\path{doi:10.1142/S0218301317500811}}.

\bibitem{Adam:2019hpq}
J.~Adam, et~al., {Observation of enhancement of charmed baryon-to-meson ratio
  in Au+Au collisions at $\sqrt{s_{NN}}$ = 200 GeV}\href
  {http://arxiv.org/abs/1910.14628} {\path{arXiv:1910.14628}}.

\bibitem{Acharya:2017kfy}
S.~Acharya, et~al., {$\Lambda_{\rm c}^+$ production in pp collisions at
  $\sqrt{s} = 7$ TeV and in p-Pb collisions at $\sqrt{s_{\rm NN}} = 5.02$ TeV},
  JHEP 04 (2018) 108.
\newblock \href {http://arxiv.org/abs/1712.09581} {\path{arXiv:1712.09581}},
  \href {http://dx.doi.org/10.1007/JHEP04(2018)108}
  {\path{doi:10.1007/JHEP04(2018)108}}.

\bibitem{Sirunyan:2019fnc}
A.~M. Sirunyan, et~al., {Production of $\Lambda_\mathrm{c}^+$ baryons in
  proton-proton and lead-lead collisions at $\sqrt{s_\mathrm{NN}}=$ 5.02
  TeV}\href {http://arxiv.org/abs/1906.03322} {\path{arXiv:1906.03322}}.

\end{thebibliography}







\end{document}